\documentclass[letterpaper,10pt]{article}

\usepackage{osameet2}

\usepackage{amsmath,amssymb}
\usepackage{subfigure}

\begin{document}

\title{Intent-Based In-flight Service Encryption in Multi-Layer Transport Networks}

\author{Mohit Chamania\textsuperscript{1,*}, Thomas Szyrkowiec\textsuperscript{1}, Michele Santuari\textsuperscript{2}, Domenico Siracusa\textsuperscript{2},\\ Achim Autenrieth\textsuperscript{1},  Victor Lopez\textsuperscript{3}, Pontus Sk\"oldstr\"om\textsuperscript{4} and St\'ephane Junique\textsuperscript{4}}
\address{\textsuperscript{1}ADVA Optical Networking, Germany \textsuperscript{2}CREATE-NET Research Center, Italy \textsuperscript{3}Telef\'onica I+D, Spain \textsuperscript{4}Acreo Swedish ICT AB, Sweden}
\email{\textsuperscript{*}email: mchamania@advaoptical.com}
\vspace{-1mm}

\begin{abstract}
We demonstrate multi-layer encrypted service provisioning via the ACINO orchestrator. ACINO combines a novel intent interface with an ONOS-based SDN orchestrator to facilitate encrypted services at IP, Ethernet and optical network layers.
\end{abstract}

%\ocis{060.4510, 060.4785}
\vspace{-3mm}

\section{Introduction}

Internet proliferation has exploded in recent years, and now offers the possibility to reach over 80\% of the population in the developed world and over 3.5 billion people worldwide \cite{prol01}.
As businesses attempt to reach this large global audience, more and more critical infrastructure is moving to the cloud, which in turn has made cyber security a growing concern.
Network encryption is an essential building block for this infrastructure, and is responsible for securing communications between trusted endpoints against interception by malicious actors.
Support for encryption is available at different network layers, and encryption at application layer is typically employed when communicating with end-users.
However, a diverse suite of protocols, most commonly employed in machine-to-machine communication, have specialized or no support for encryption, complicating the adoption of network encryption.
In-flight encryption is a standard mechanism to address this by encrypting traffic between trusted \emph{sites} at the media layer (IP/MAC/PHY).

The demand for encrypted service delivery introduces additional complexity to a service request made to a carrier network.
Service provisioning in the current ecosystem involves the manual translation of high-level requirements by an application into technology-specific configurations which are then applied on the underlying infrastructure.
The manual intervention makes the complete service provisioning procedure slow and error-prone, and intent-based API paradigms have emerged as a novel mechanism to automate this procedure \cite{dismi}.
Intent APIs define high-level primitives that can be used by an application to define its requirements from the underlying infrastructure, and \emph{intent compilers} are responsible for identifying potential service configurations based on the capabilities of the infrastructure at hand.

We demonstrate an intent-based in-flight encryption delivery mechanism in a multi-layer network infrastructure.\footnote{This research has received funding from the European Commission within the H2020 Programme, ACINO project, Grant Number 645127.}
In the demonstration, high-level intents, defined by the ACINO project \cite{dismi}, will be extended to indicate requirements on encryption.
These definitions will be processed by an SDN orchestrator \cite{acino}, which will compute and provision a service in a multi-layer network.

\vspace{-3mm}

\section {Architecture}

\begin{figure*}[t]
    \centering
	\subfigure[The ACINO Orchestrator Architecture]{
		\includegraphics[width=0.28\textwidth] {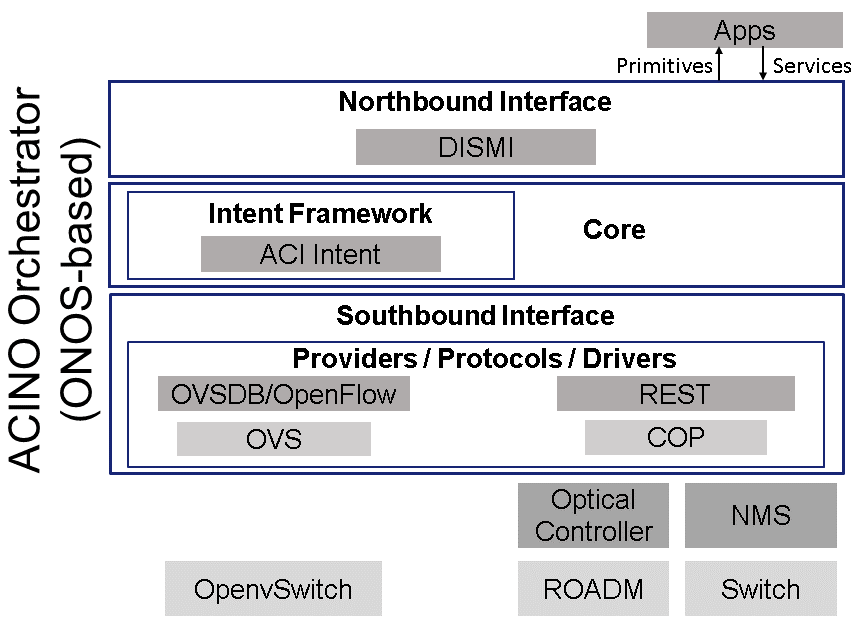}
		\label{fig1a}
	}\hspace{5mm}
	\subfigure[Experimental Testbed Setup]{
		\includegraphics[width=0.45\textwidth] {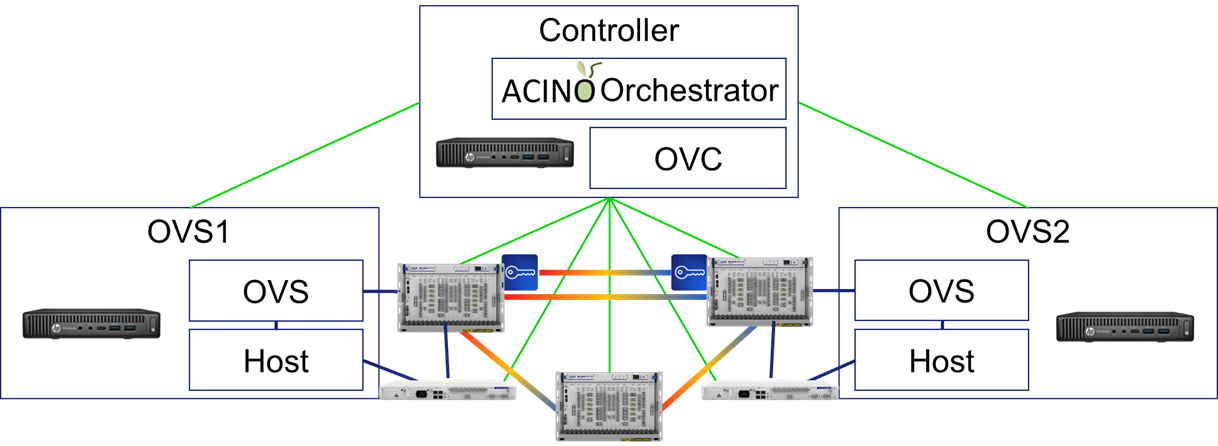}
		\label{fig1b}
	}
\caption {Architecture and Testbed Setup}
\vspace{-8mm}
\end{figure*}

The ACINO orchestrator, presented in Fig. \ref{fig1a}, is a logically centralized SDN controller based on ONOS \cite{onos}, with the ability to translate technology-agnostic application service requirements into configuration requests for the underlying packet (IP/OpenFlow) and optical networks.

The Dynamic Intent-driven Service Management Interface (DISMI) is a network independent and requirement-aware intent API exposed to end-user applications.
It exposes the available features of the orchestrator as primitives in a technology agnostic fashion, and applications can make service requests based on these primitives.
For example, technology agnostic endpoint primitives are used to identify \emph{sites}, and encryption primitives are provided for applications with strict regulatory requirements on encryption, like banking.
After validating the service request, the DISMI translates it into one or multiple network-aware intents called Application Centric Intents (ACI) that are finally compiled by the Intent Framework into installable network operations.
The ACI and its compiler are developed as an enhancement of the ONOS Intent Framework to jointly manage IP and optical layers.
In particular, the compiler is responsible for evaluating and translating ACIs into network service requests and constraints and generate the installable multi-layer operations.
The southbound interface takes care of the intent installation by translating the installable operations into protocol procedures sent to the network.
We leverage OpenFlow, OVSDB and YANG descriptions, e.g. Control Orchestration Protocol \cite{cop} in combination with HTTP as a REST protocol and extensions to support encryption when available.

\vspace{-3mm}
\section{Testbed Overview and Requirements}

Fig. \ref{fig1b} outlines the testbed used to demonstrate multi-layer in-flight encryption.
The optical network consists of a three node ring with ADVA FSP3000 ROADMs, two of which are equipped with transponders supporting optical encryption.
Two MACsec capable hardware switches are connected to regular transponders on the optical network to demonstrate encryption on the Ethernet layer.
Two PCs running Open vSwitch (OVS) instances are also connected to the WDM ring to demonstrate IP encryption using IPsec enabled GRE tunnels.
An additional machine hosts the controllers, i.e. the ACINO orchestrator and the Optical Virtualization Controller (OVC), is connected to all network entities through an out-band channel.
The hardware and software will be deployed in a lab at ADVA premises, and the operations will be performed via a Web interface from the venue.
To access the lab an Internet connection, that allows the creation of a VPN tunnel, is required.
\vspace{-3mm}
\section {Innovation}
%Applications need a mechanism to describe requirements directed at an underlying network infrastructure in a technology-agnostic manner.
Client applications lack a mechanism to define their needs in a technology agnostic manner.
%These requirements can include familiar networking constraints (bandwidth, delay etc.) but may also contain \emph{soft} requirements, which in turn poses restrictions on the underlying services. 
The in-flight encryption service we propose to demonstrate is a prime example of how standard networking and additional \emph{soft} requirements like regulatory compliance can be effectively communicated from an application to the underlying infrastructure.
For example, in-flight encryption for govermental applications in Germany must comply with BSI standards, while healthcare applications in the US must comply with HIPAA.
We demonstrate how the ACINO architecture uses an intent-based interface (DISMI) to receive technology-agnostic requests with specific requirements and effectively convert them to network service requests, which are finally compiled into installable services on the multi-layer network.
We also demonstrate the provisioning of encrypted services using SDN protocols (with minor extensions) on the IP, Ethernet and Optical network layers.

\vspace{-3mm}

\section {Relevance}
This work addresses a key requirement to support interactions between applications and transport networks. Applications cannot treat the network as a black box, and need a mechanism to describe requirements directed at an underlying network infrastructure in a technology-agnostic manner. The work also builds upon established open-source SDN frameworks to incorporate application awareness in service orchestration. The project is also actively contributing to the ONOS community to improve dissemination of the project.


\vspace{-2mm}

\begin{thebibliography}{99}

\bibitem{prol01} Internet live stats. [Online]. Available: http://www.internetlivestats.com/internet-users/

\bibitem{dismi} P.~Sk\"oldstr\"om and S.~Junique, ``Application-centric networks and the future 5G transport,'' in \emph{17th International Conference on Transparent Optical Networks (ICTON)}, July 2015.

\bibitem{acino} T.~Szyrkowiec et al.,
``First demonstration of an automatic multilayer intent-based secure service creation by an open source sdn orchestrator,''
in \emph{42nd European Conference on Optical Communication (ECOC)}, Sep. 2016.

\bibitem{onos} P. Berde et al., ``ONOS: Towards an Open, Distributed SDN OS", in HotSDN, 2014.
\bibitem{cop} R. Muñoz et al., ``Transport Network Orchestration for End-to-End Multi-layer Provisioning Across Heterogeneous SDN/OpenFlow and GMPLS/PCE Control Domains'', in Journal of Lightwave Technology, April 2015, Vol. 33, No. 8, pp. 1540 - 1548. 
\end{thebibliography}
\end{document}